\documentclass[aps,twocolumn,superscriptaddress,showkeys,showpacs]{revtex4}
\usepackage{amssymb}
\usepackage{amsfonts}
\usepackage{amsmath}
\usepackage{amsthm}
\usepackage{epsfig}
\usepackage{color} 
\usepackage{subfigure}
\usepackage{bbm}

\newcommand{\eps}{\varepsilon}

\def\mean#1{\langle#1\rangle}
\newcommand{\Ueff}{U_{\text{eff}}}
\newcommand{\Deff}{D_{\text{eff}}}

\begin{document}
\title{Noise-induced transitions in a double-well oscillator with nonlinear dissipation}

\author{Vladimir V. Semenov}
\email[corresponding author: ]{semenov.v.v.ssu@gmail.com}
\affiliation{Department of Physics, Saratov State University, Astrakhanskaya str., 83, 410012, Saratov, Russia}
\author{Alexander B. Neiman}
\email{neimana@ohio.edu}
\affiliation{Department of Physics and Astronomy, Ohio University, Athens, Ohio 45701, USA}
\author{Tatyana E. Vadivasova}
\affiliation{Department of Physics, Saratov State University, Astrakhanskaya str., 83, 410012, Saratov, Russia}
\author{Vadim S. Anishchenko}
\affiliation{Department of Physics, Saratov State University, Astrakhanskaya str., 83, 410012, Saratov, Russia}

\date{\today}

\begin{abstract}
We develop a model of bistable oscillator with nonlinear dissipation.  Using a numerical simulation and an electronic circuit realization of this system we study its response to additive noise excitations. We show that depending on noise intensity the system undergoes multiple qualitative changes in the structure of its steady-state probability density function (PDF). In particular, the PDF exhibits two pitchfork bifurcations versus noise intensity, which we describe using an effective potential and corresponding normal form of the bifurcation. These stochastic effects are explained by the partition of the phase space by the nullclines of the deterministic oscillator.
\end{abstract}
\maketitle

\section{Introduction}
Bistable dynamics is typical for many natural systems  in physics \cite{gibbs1985,hanggi1990, risken1996}, chemistry \cite{kramers1940,schloegl1972}, biology \cite{goldbetter1997,guidi1998,ozbudak2004,shilnikov2005,smits2006,benzi2010}, ecology \cite{may1977,guttal2007}, geophysics \cite{benzi1981,benzi1982,nicolis1982}. 
The simplest kind of bistability occurs when a system possesses two stable equilibria in the phase space, separated by a saddle.
Adding noise gives rise to random switchings between the deterministically stable states, resulting in a steady state probability density with two local maxima. The Kramers oscillator is a classical example of the stochastic bistable system describing Brownian motion in a double-well potential \cite{hanggi1990,kramers1940,freund2003},
\begin{equation}
\label{general}
\dot{y}=v, \quad \dot{v} = -\gamma v - \frac{dU(y)}{dy} + \sqrt{2\gamma D}\,n(t),
\end{equation}
where $\gamma$ is the (constant) drag  coefficient, $U(y)$ is a double-well potential and $n(t)$ is Gaussian white noise, $D$ is the noise intensity.
Two-dimensional equilibrium PDF is: 
\begin{equation}
\label{P(y,v)}
 P(y,v)=k \exp \left[ -\frac{1}{D} \left( \frac{v^2}{2}+U(y) \right) \right],
 \end{equation}
with the normalization constant $k$, and possess two maxima, corresponding to the potential wells, separated by a saddle point of the potential. This structure does not depend on the noise intensity $D$: although the peaks in the PDF are smeared out, their position is invariant with respect to increase of noise intensity. 

External random perturbations may result in the so-called noise-induced transitions whereby stationary PDF changes its structural shape, e.g. number of extrema, when noise intensity varies \cite{horsthemke1984}. 
Such transitions  may occur both with multiplicative noise as in the original Horsthemke-Lefever scenario \cite{horsthemke1984}, and with additive noise (see e.g. \cite{Lutz1985,zakharova2010}). For example, with a multiplicative noise stochastic bistable oscillator shows reentrant (multiple) noise-induced transitions when the noise intensity varies \cite{mallick2004}. Noise-induced transitions were observed in excitable systems ranging from a single excitable neurons \cite{tanabe2001noise,kromer2014,neiman2007} to coupled excitable elements and media \cite{ullner2003noise,zaks2005noise}.
In many cases noise-induced transitions are not true bifurcations \cite{toral2011noise},  rather they underlie qualitative changes of stochastic dynamics when noise strength is the control parameter. Noise-induced {\it phase} transitions were studied in spatially distributed systems perturbed by multiplicative noise \cite{van1994noise} and were shown to exist for the case of additive noise \cite{zaikin1999nonequilibrium}. 
In this paper we develop a generalized bistable oscillator with nonlinear dissipation and report on a multiple noise-induced transitions due to additive noise in this system. We first develop an electronic circuit of the oscillator and demonstrate noise-induced transitions in analog experiment. Second, we use a corresponding deterministic model of the circuit and numerical simulations of stochastic model to explain mechanisms of noise-induced transitions.

\section{Model and Methods}
\begin{figure}[t]
\centering
\includegraphics[width=0.4\textwidth]{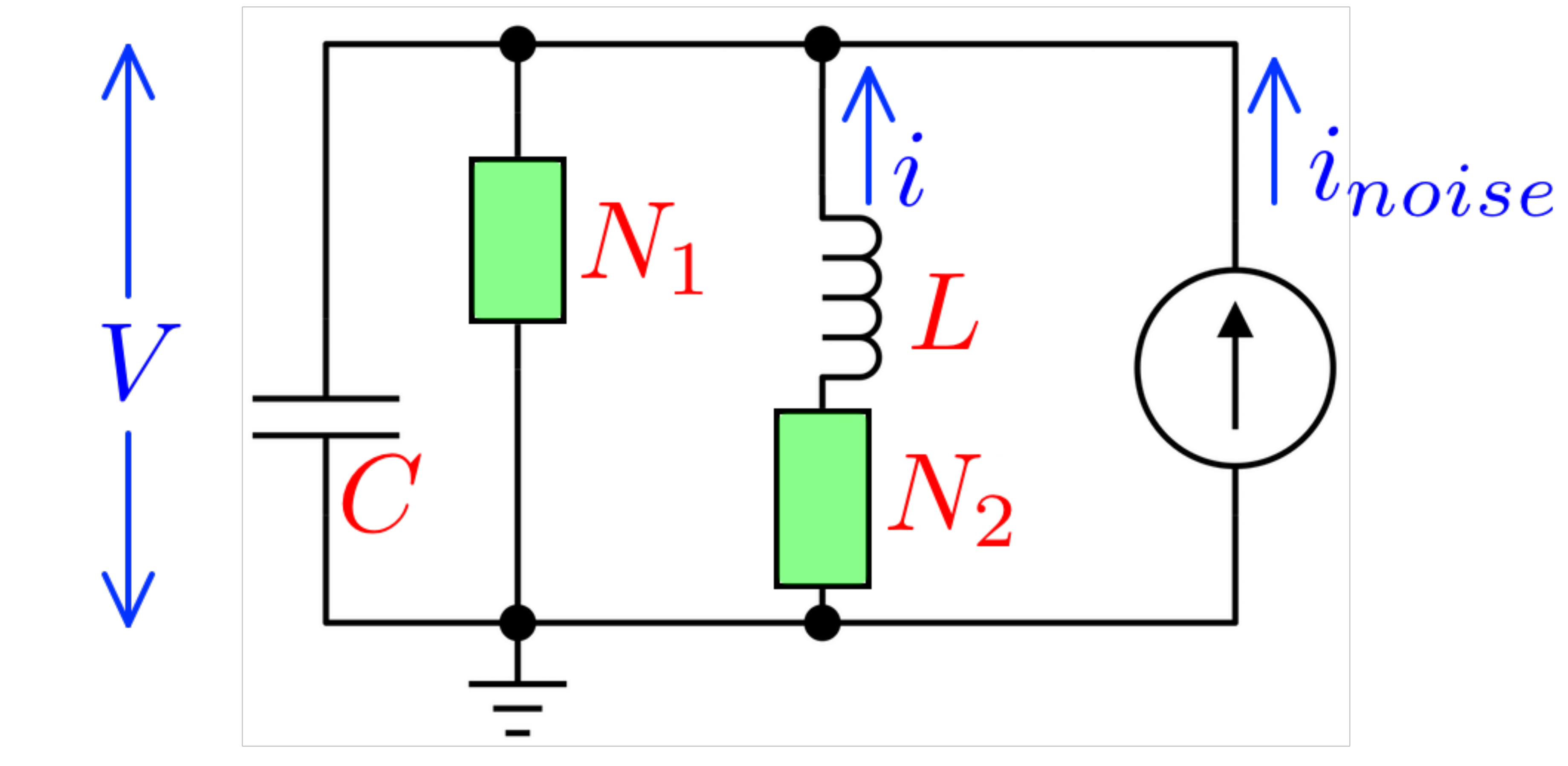} 
\caption{Circuit diagram of the model.}
\label{fig1}
\end{figure}

Fig.~\ref{fig1} shows a circuit diagram with two nonlinear elements N1 and N2 with the S- and N-type of the I-V characteristic, respectively: $i_\mathrm{N1} = F(V)$, $V_\mathrm{N2} = G(i)$. The circuit is similar to Nagumo's tunnel diode neuron model \cite{nagumo1962active,izhikevich2006fitzhugh}, except it contains nonlinear resistor N2 in series with the inductor, L.
The circuit also includes a source of broadband Gaussian noise current $i_\mathrm{noise}(t)$, which will be assumed white in the following.
By using the Kirchhoff's current law the following differential equations for the voltage $V$ across the capacitance $C$  and the current $i$ through the inductance $L$ can be derived:
\begin{equation}
\label{eq1}
\left\lbrace
\begin{array}{l}
C\dfrac{dV}{dt'}+i_\mathrm{N1} + i + i_{noise}(t') = 0, \\
V = L\dfrac{di}{dt'}+V_\mathrm{N2}. \\
\end{array}
\right.
\end{equation}

In the dimensionless variables $x=V / v_{0}$ and $y=i/ i_{0}$ with $v_{0}= 1$~V, $i_{0}=1$~A and dimensionless time $t=[(v_{0}/(i_{0}L)]t'$, Eq.(\ref{eq1}) can be re-written as,
\begin{equation}
\label{initial}
\left\lbrace
\begin{array}{l}
\eps \dot{x} = -y-F(x) - \sqrt{2D}\,n(t), \\
\dot{y}=x-G(y), \\
\end{array}
\right.
\end{equation}
%
The parameter $\eps$ sets separation of slow and fast variables of the system, $\varepsilon=(C/L)(v_0/i_0)^2$. The first equation for the voltage contains an additive source of white Gaussian noise, $n(t)$, with the intensity $D$: $\mean{n(t)}=0$, $\mean{n(t)n(t+\tau)}=2D\delta(\tau)$.
Depending on the shape of the functions $F(x)$ and $G(y)$ the circuit demonstrates wide range of dynamics, including various types of bistability, self-sustained oscillations and excitability. It allows to observe a wide range of dynamical regimes: from the behavior like in oscillator (\ref{general}) with a double-well potential to dynamics of an excitable oscillator or a bistable self-sustained oscillators. This letter is restricted to bistability of two stable equilibria with N-shape function $G(y)=-ay+by^3$ with positive coefficients $a$ and $b$.  

We start with  a linear resistor N1, $F(x)=c_1 x$, with positive $c_1$. The circuit is described by,
\begin{equation}
\label{linear}
\begin{array}{l}
\eps \ddot{y} = -((3by^{2}-a) \eps + c_1) \dot{y}-\\
(1-c_1 a+c_1 b y^2) y - \sqrt{2 D} n(t).\\
\end{array}
\end{equation}
The friction is nonlinear, but depends solely on the "coordinate" variable, $y$. For sufficiently small $\eps$, $c_1 \gg \eps (3by^2-a)$ and dissipation becomes essentially linear. Then the system is closely akin to Kramers oscillator, Eq.(\ref{general}), with  {\it linear} friction. Our analog and numerical simulations indicated no qualitative differences in dynamics of the Kramers oscillator and the circuit (\ref{linear}) with linear resistor N1, i.e. no noise-induced qualitative change in the stationary PDF. 

Next, we consider  the case of nonlinear resistor N1 with $F(x)=c_{1}x-c_{3}x^{3}+c_{5}x^{5}$ and with fixed positive coefficients $c_{1}=1,c_{3}=9,c_{5}=22$. A variety of the electronic elements and circuits have I-V characteristic like that. For example, N1 can be realized by  the so-called lambda-diode  circuit \cite{kano1975}. In this case the following equations describes the system under study,

\begin{equation}
\label{system}
\left\lbrace
\begin{array}{l}
\varepsilon \dot{x} = -y-c_{1}x+c_{3}x^3-c_{5}x^5-\sqrt{2D}n(t), \\
\dot{y}=x+ay-by^3. \\
\end{array}
\right.
\end{equation}

Eqs. (\ref{system}) can be written in the "coordinate-velocity" form with dynamical variables $y$, $v\equiv \dot{y}$ and $x=v-ay+by^3$:
\begin{equation}
\label{system2}
\left\lbrace
\begin{array}{l}
\dot{y}=v, \\
\eps \dot{v} =-y-c_1 (v-ay+by^3)+c_3 (v-ay+by^3)^3-\\ 
c_5 (v-ay+by^3)^5+ \eps v ( a -3 by^2)-\sqrt{2D}n(t).
\end{array}
\right.
\end{equation}
In the oscillatory form (\ref{system2}) becomes,

\begin{equation}
\label{system3}
\ddot{y}+q_{1}(y,\dot{y})\dot{y}+\frac{1}{\varepsilon}q_{2}(y)=-\sqrt{2 D}n(t),
\end{equation}
where $q_{2} (y)$ defines the form of the potential, and $q_{1}(y, \dot{y})$ is the nonlinear dissipation,
\begin{equation}
\label{q1q2}
\left.
\begin{array}{l}
q_{1}(y,\dot{y})=-a+3by^{2}+ \dfrac{1}{\varepsilon}(c_{1}-\\
c_{3} \sum\limits_{n=1}^{3} \frac{3!}{n!(3-n)!} \dot{y}^{n-1} (by^{2}-a)^{3-n}y^{3-n}+ \\
+c_{5} \sum\limits_{n=1}^{5} \frac{5!}{n!(5-n)!} \dot{y}^{n-1} (by^{2}-a)^{5-n}y^{5-n}), \\
 \\
q_{2}(y)= y+c_{1}(by^{2} -a)y-\\
\\
c_{3}(by^{2} -a)^{3} y^{3} +c_{5}(by^{2} -a)^{5} y^{5}. \\
\end{array}
\right.
\end{equation}
We note that unlike for the case of linear resistor Eq.(\ref{linear}), the system's dissipation  now depends on both coordinate and velocity, $y$ and $\dot{y}$. As we show below, it gives the main reason of the qualitative difference between dynamics of the system (\ref{system2}) and behavior of the models (\ref{general}) and (\ref{linear}).

The proposed system  was studied by means of analog and numerical simulations.
Experimental electronic setup was developed by using principles of analog modeling of stochastic systems \cite{moss1989,luchinsky1998}. The main part of the analog model is the operational amplifier integrator, whose output voltage is proportional to the input voltage integrated over time: $V_{out}=-\dfrac{1}{R_{0}C_{0}} \int\limits_0^t V_{in}dt\quad$ or  $\quad R_{0}C_{0}\dot{V}_{out}=- V_{in} $.
%
\begin{figure}[t]
\centering
\includegraphics[width=0.5\textwidth]{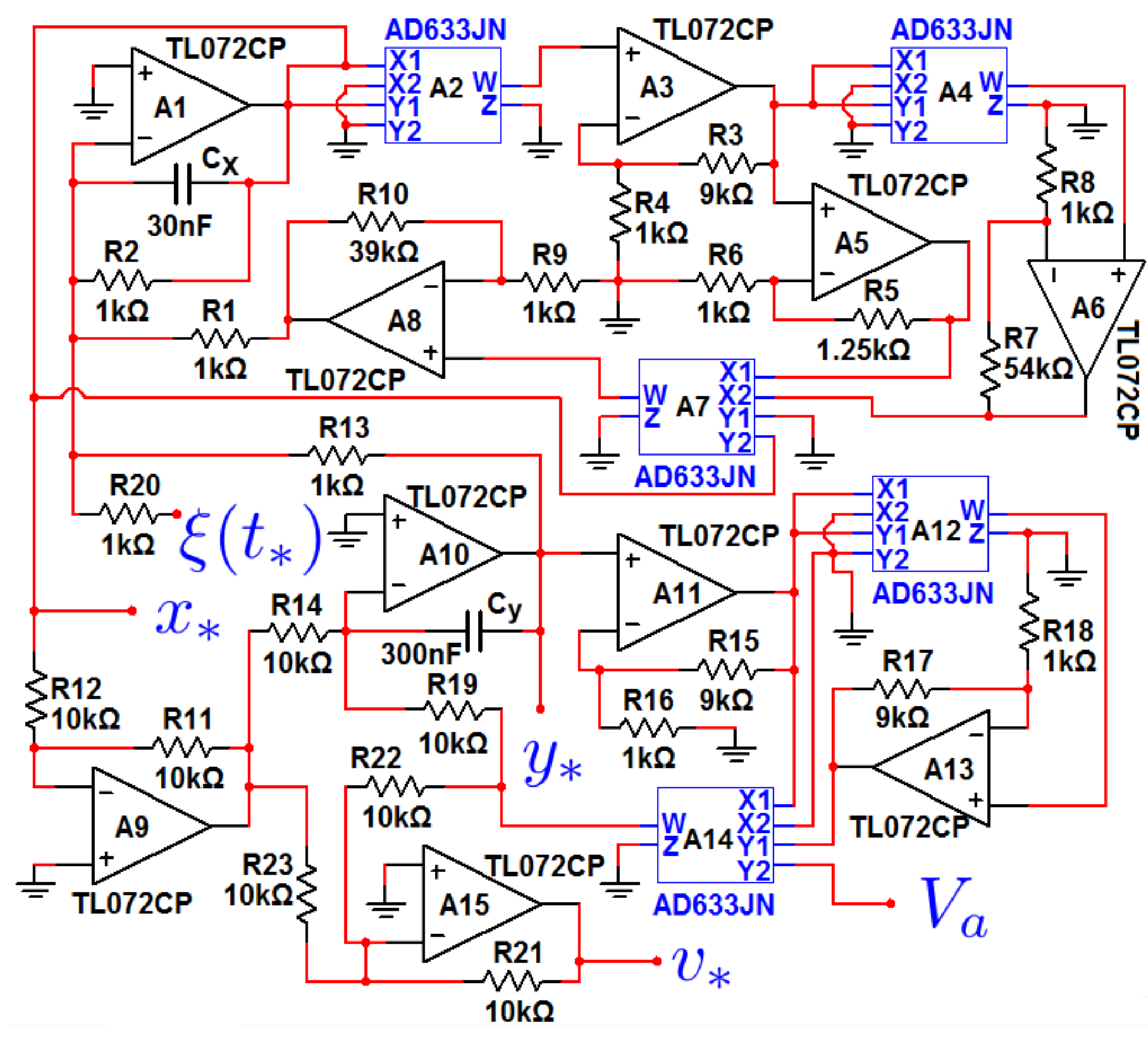} \\
\caption{Scheme of the experimental setup.}
\label{fig2}
\end{figure}  
%
Circuit diagram is shown in Fig. \ref{fig2}. It contains two integrators, A1 and A10, whose output voltages are taken as the dynamical variables, $x_{*}$ and $y_{*}$, respectively. 
Then the signals $x_{*}$ and $y_{*}$ are transformed in order to match expressions of the right hand side of Eqs.~(\ref{system}). The necessary signal transformations are carried out by using analog multipliers AD633JN and the operational amplifiers TL072CP connected in the inverting and non-inverting amplifier configurations. Finally, transformed signals come to the input of the integrators as $V_{in}$. The experimental setup allows to obtain the instantaneous values of the variables $x_{*},y_{*}$ and $v_{*}=\dot{y}_{*}=x_{*}+ay_{*}-by_{*}^{3}$. Time series were recorded from corresponding outputs
(marked in Fig.~\ref{fig2}) using an acquisition board (National Instruments NI-PCI 6133). All signals were digitized at the sampling frequency of 50 kHz. 
150~s long realizations were used for further offline processing. A noise generator G2-59 was used to produce broadband Gaussian noise, whose spectral density was almost constant in the frequency range 0  -- 100 kHz. In this frequency range noise can be approximated by white Gaussian.

The circuit in Fig. \ref{fig2} is described by the following equations:
\begin{equation}
\label{exp}
\left\lbrace
\begin{array}{l}
R_{x}C_{x} \dfrac{dx_{*}}{dt_{*}} = -y_{*}-c_{1}x_{*}+c_{3}x_{*}^3-\\
c_{5}x_{*}^5-\xi(t_{*}),\\
R_{y}C_{y}\dot{y}_{*}=x_{*}+ay_{*}-by_{*}^3, \\
\end{array}
\right.
\end{equation}
where $C_{x}=30$~nF, $C_{y}=300$~nF,  $R_{x}=$1~K$\Omega$ is the resistance at the integrator A1 ($R_{1}=R_{2}=R_{13}=R_{20}=R_{x}=1~$K$\Omega$), $R_{y}=10$~K$\Omega$ is the resistance at the integrator A10 ($R_{14}=R_{19}=R_{y}=10$~K$\Omega$). The parameter $a$ is equal to the input value of the voltage $V_{a}$ at the analog multiplier A14, $b=10\left(1+\dfrac{R_{17}}{R_{18}}\right)$;  $c_{1}=1$, $c_{3}=4\left(1+\dfrac{R_{5}}{R_{6}}\right)$, $c_{5}=0.4\dfrac{R_{7}}{R_{8}}$;
$\varepsilon=\dfrac{R_{x}C_{x}}{R_{y}C_{y}}$. Transition to dimensionless equations (\ref{system2}) is then carried out by 
substitution $t=t_{*}/\tau_0$, $x=x_{*}/v_{0}$, $y=y_{*}/v_{0}, v=v_{*}/v_{0}$, where $\tau_0=R_{0}C_{0}=R_{y}C_{y}=3$~ms is the circuit's time constant  and $v_{0}=1$~V. The intensity $D'$ of noise generated by noise generator is related to the dimensionless $D$ of Eq.(\ref{system2}) by $D = D'/ \tau_0$.

Numerical simulations were carried out by the integration of Eq. (\ref{system2}) using the Heun method \cite{manella2002} with time step $\Delta t=0.0001$.

In the following the parameters of deterministic system were set to $\varepsilon=0.01$, $a=1.2$, $b=100$, $c_{1}=1$, $c_{3}=9$, $c_{5}=22$. For this set of parameters 
the deterministic system possesses three equilibria: two stable nodes and a saddle at the origin. Noise intensity, $D$, was used as a control parameter in analog experiments and numerical simulations.
 
\begin{figure}[t]
\centering
\includegraphics[width=0.5\textwidth]{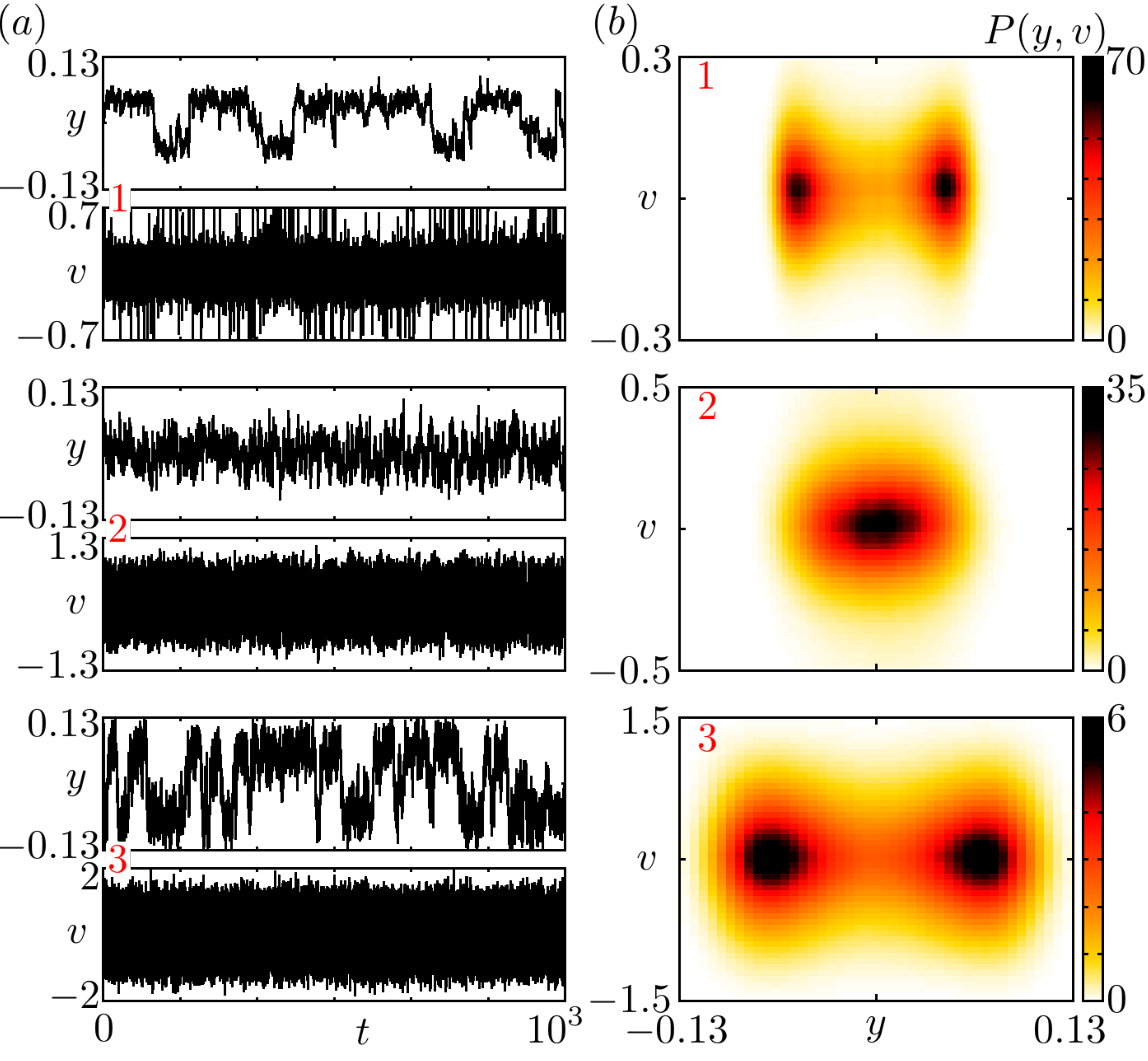} 
\caption{Noise-induced transitions in analog simulations. 
(a): Time traces of state variables for various values of noise intensity:
1 -- $D=1.51\times 10^{-4}$, 2 -- $D=3.78\times 10^{-4}$, 3 -- $D=3.00\times 10^{-3}$.
(b): Stationary probability density functions (PDF) $P(y,v)$ corresponding to traces on (a).
Other parameters are: $\eps=0.01, c_{1}=1, c_{3}=9, c_{5}=22, a=1.2, b=100$.}
\label{fig3}
\end{figure}  

\section{Noise-induced transitions and effective potential}
\subsection{Analog Experiment}
Experiments with the analog circuit showed that noise strength is true control parameter of the system.
For weak noise the circuit exhibits bistable dynamics with a typical hopping between two metastable states and two peaks in the PDF shown in Fig.~\ref{fig3}(a1,b1).
Increase of noise intensity leads to qualitative change in the stochastic dynamics: switching between two states disappears and so the PDF has a single global maximum [Fig.~\ref{fig3}(a2,b2)]. Furthermore, larger noise results in yet another noise-induced transition whereby dynamics becomes again bistable  with two-state hopping and double-peaked stationary PDF [Fig.~\ref{fig3}(a3,b3)].
\begin{figure}[t!]
\centering
\includegraphics[width=0.5\textwidth]{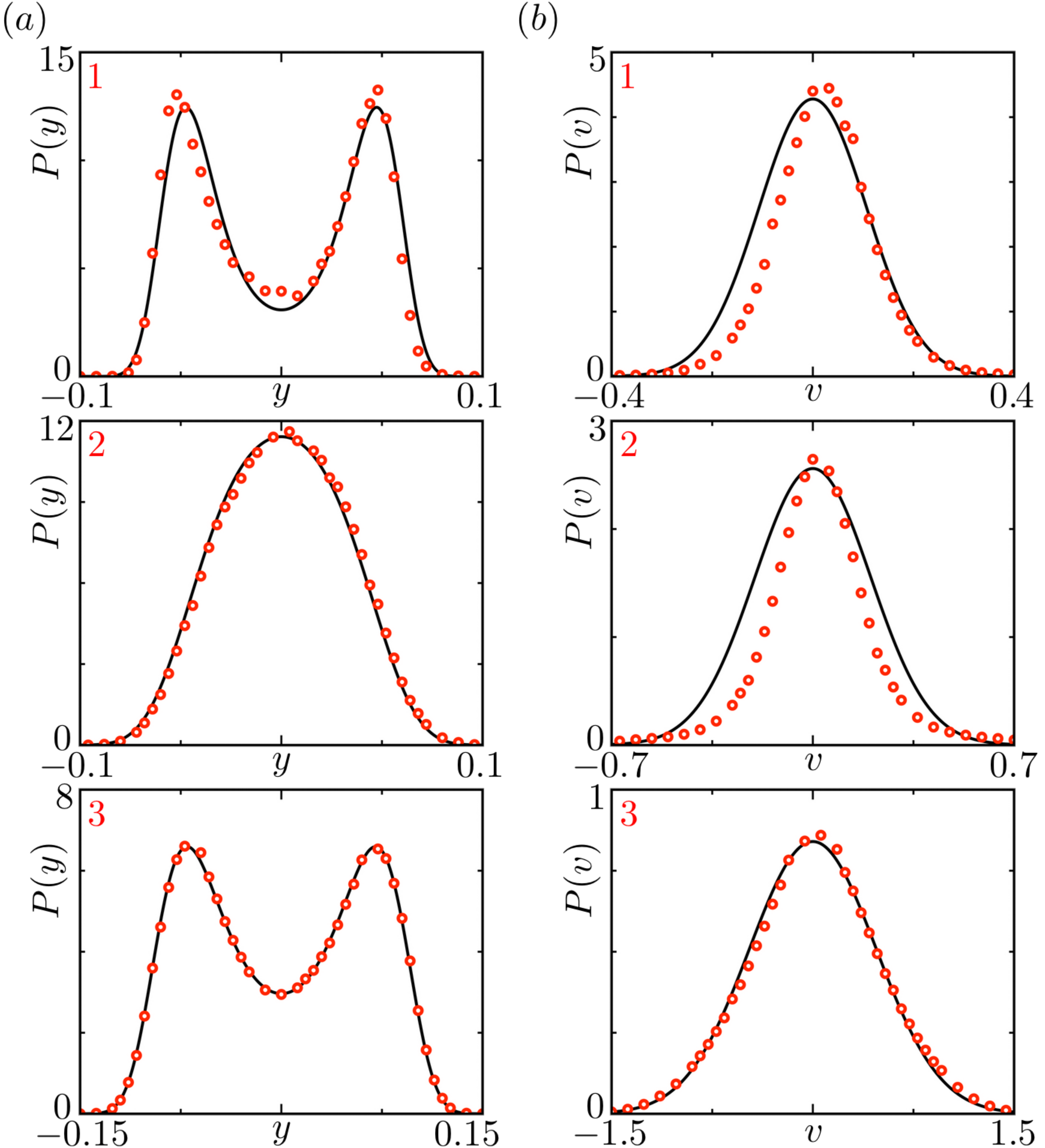} 
\caption{Noise-induced bifurcations in analog simulations. (a): Marginal PDF of the coordinate, $P(y)$, (red circles) and its fit using the effective potential Eq.(\ref{pdffit}) (lines). (b): Marginal velocity PDF, $P(v)$, (red circles) and a Gaussian distribution with the same mean and variance $\sigma^{2}=\Deff$ (lines). 
Noise intensity, parameters of the effective potential and the effective noise intensity are: 1 -- $D=1.51\times 10^{-4}$, $\alpha=14.35$, $\beta=3193.5$, $\Deff=1.15\times 10^{-2}$; 2 -- $D=3.78\times 10^{-4}$, $\alpha=-9.71$, $\beta=2532.4$, $\Deff=4.07\times 10^{-2}$; 3 -- $D=3.00\times 10^{-3}$, $\alpha=76.12$, $\beta=7707.1$, $\Deff=2.35 \times 10^{-1}$. Other parameters are the same as in the previous figure.}
\label{fig4}
\end{figure}  
%
These multiple noise-induced transitions are shown in Fig.~\ref{fig4} for the marginal PDFs of the coordinate and velocity variables,
$P(y)=\int_{-\infty}^\infty P(y,v) dv$, $P(v)=\int_{-\infty}^\infty P(y,v) dy$. While the PDF of the coordinate, $P(y)$, shows noise-controlled changes of its modality from bimodal to unimodal and back to bimodal, the velocity PDF shows no modality change. In this respect, the velocity distribution is similar to
one of the Kramers oscillator. Although,  indeed we do not expect $P(v)$ to follow a Gaussian distribution as for the Kramers oscillator.

\begin{figure}[t!]
\centering
\includegraphics[width=0.5\textwidth]{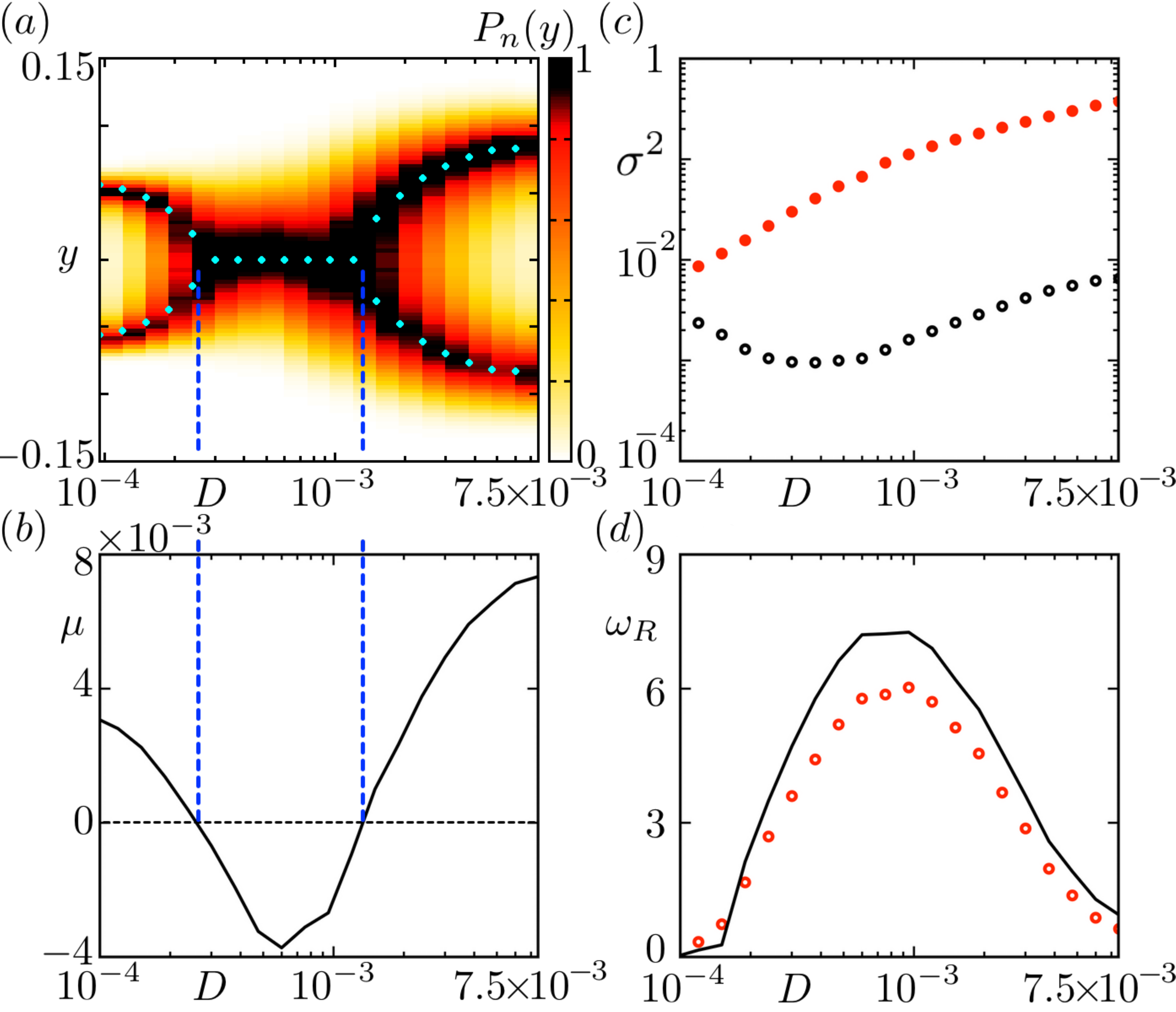} 
\caption{Noise-induced transitions in analog simulations. (a): Marginal coordinate PDF vs noise intensity, $P(y;D)$. 
For each value of noise intensity, $D$, the PDF $P(y)$, was normalized to its maximal value, i.e. $P_n(y) = P(y)/P_\mathrm{max}$. Blue dots show minima of the corresponding effective potential (\ref{pdffit}).
(b): Effective bifurcation parameter $\mu(D)$. Vertical dashed lines indicate positions of two pitchfork bifurcations. 
(c): Coordinate and velocity variance vs noise intensity. The coordinate variance, $\sigma^2_y$ is shown by the open black circles; the velocity variance which equals the effective noise intensity, $\Deff=\sigma^2_v$, is shown by filled red circles. 
(d): Rice frequency, $\omega_R$, vs noise intensity. Open red circles show results of analog simulations; solid line shows
Eq.(\ref{rise}) with the effective potential $\Ueff$ and effective noise $\Deff$.} 
\label{fig5}
\end{figure}

Noise-induced transitions are most apparent in a diagram of the marginal PDF of coordinate, $P(y)$, plotted vs noise intensity, $D$, in Fig.~\ref{fig5}(a). 
Here the PDF $P(y;D)$ is shown as filled contour lines allowing to track positions of PDF's extrema vs noise intensity and clearly indicates  two pitchfork bifurcations. 

The described stochastic dynamics can be represented in terms of an effective potential $\Ueff(y)$, such that  the marginal coordinate PDF  is
$P(y) \propto \exp[-\Ueff(y)/\Deff]$, with an effective noise intensity, $\Deff$.
Fig.~\ref{fig4}(a) shows that the marginal coordinate PDF, $P(y)$, can be nicely fitted by 
\begin{equation}
\label{pdffit}
P(y) = k_1 \exp\left[-\frac{1}{\Deff}\Ueff(y)\right],\,\,
\Ueff(y) = -\alpha y^2+\beta y^4,
\end{equation}
where $k_1$ is the normalization constant. The effective noise intensity is calculated as the variance of velocity, $\Deff(D) \equiv \mathrm{var}[v]$, while the parameters $\alpha$, $\beta$ are estimated from the least square fit of the experimentally measured marginal PDF, $P(y)$.
The effective potential can be re-written in the form, $\Ueff(y)= 4\beta( -\mu \, y^2/2 +y^4/4)$, with $\mu=\alpha/(2\beta)$. The shape of the effective potential is determined by the effective bifurcation parameter, $\mu(D)$. 
Thus, noise-induced transitions of the circuit can be effectively described by the normal form of the pitchfork bifurcation perturbed by white noise, \cite{meunier1988noise},
\begin{equation}
\dot{y} = 4\beta( \mu y - y^3) +\sqrt{2\Deff}\,\xi (t).
 \label{normal.eq}
\end{equation}
With positive $\beta$, the dynamics of Eq.(\ref{normal.eq}) is bistable for $\mu(D) >0$, monostable for $\mu(D) <0$ and critical at $\mu(D)=0$. Fig.~\ref{fig5}(b) shows the dependence of the effective bifurcation parameter vs noise intensity, $\mu(D)$, and clearly indicates two pitchfork bifurcations at $D=2.6 \times 10^{-4}$ and $D=1.34\times 10^{-3}$ which match the bifurcation diagram,  Fig.~\ref{fig5}(a), obtained from the marginal PDF, $P(y;D)$. Furthermore, as indicated in Fig.~\ref{fig5}(c), while the velocity variance, $\sigma^2_v=\mathrm{var}[v]$ (and so the effective noise intensity, $\Deff$) increases monotonously with $D$, the coordinate variance,  $\sigma^2_y=\mathrm{var}[y]$, shows a non-monotonous dependence, reflecting transitions from bistable to monostable regimes. 

Stochastic bistable oscillators are characterized by two time scales: fast intrawell fluctuations and slower interwell switching. 
The mean frequency of bistable oscillator can be quantified by the Rice frequency \cite{callenbach2002oscillatory,freund2003},
which is the rate of zero-crossings by the oscillator's coordinate with positive velocity, 
$\omega_R = 2\pi \int_0^\infty v P(y=0,v) dv$. For the Kramers oscillator (\ref{general}) with linear friction the Rice frequency reads
\cite{callenbach2002oscillatory}, 
\begin{equation}
\label{rise}
\omega_{R}=\frac{\sqrt{2\pi D} \exp \left[ -\frac{U(0)}{D} \right] }{\int\limits_{-\infty}^{\infty}\exp \left[-\frac{U(y)}{D}\right] dy}.
\end{equation}
It increases with noise intensity \cite{callenbach2002oscillatory}, reflecting the increase of the Kramers rate of transitions between metastable states: the longer is the residence in metastable states, the smaller is the Kramers's rate and the Rice frequency. The dependence of the Rice frequency on noise intensity for our circuit is non-monotonous [Fig.~\ref{fig5}(d)]: $\omega_R$ is low for weak and strong noise, where the system is bistable, and attains its maximal value for intermediate noise, corresponding to effective monostable dynamics with the minimal value of the bifurcation parameter, $\mu$. 
Unlike for the Kramers oscillator (\ref{general}) which is characterized by Gaussian velocity distribution, the velocity PDF, $P(v)$, for our bistable oscillator is non-Gaussian [Fig.~\ref{fig4}(b)]. Nevertheless the non-monotonous dependence $\omega_R(D)$ is qualitatively approximated by Eq.(\ref{rise}) with the effective potential and noise intensity, i.e. with the substitution $\Ueff \to U$, $\Deff \to D$ in (\ref{rise}).
\begin{figure}[t!]
\centering
\includegraphics[width=0.5\textwidth]{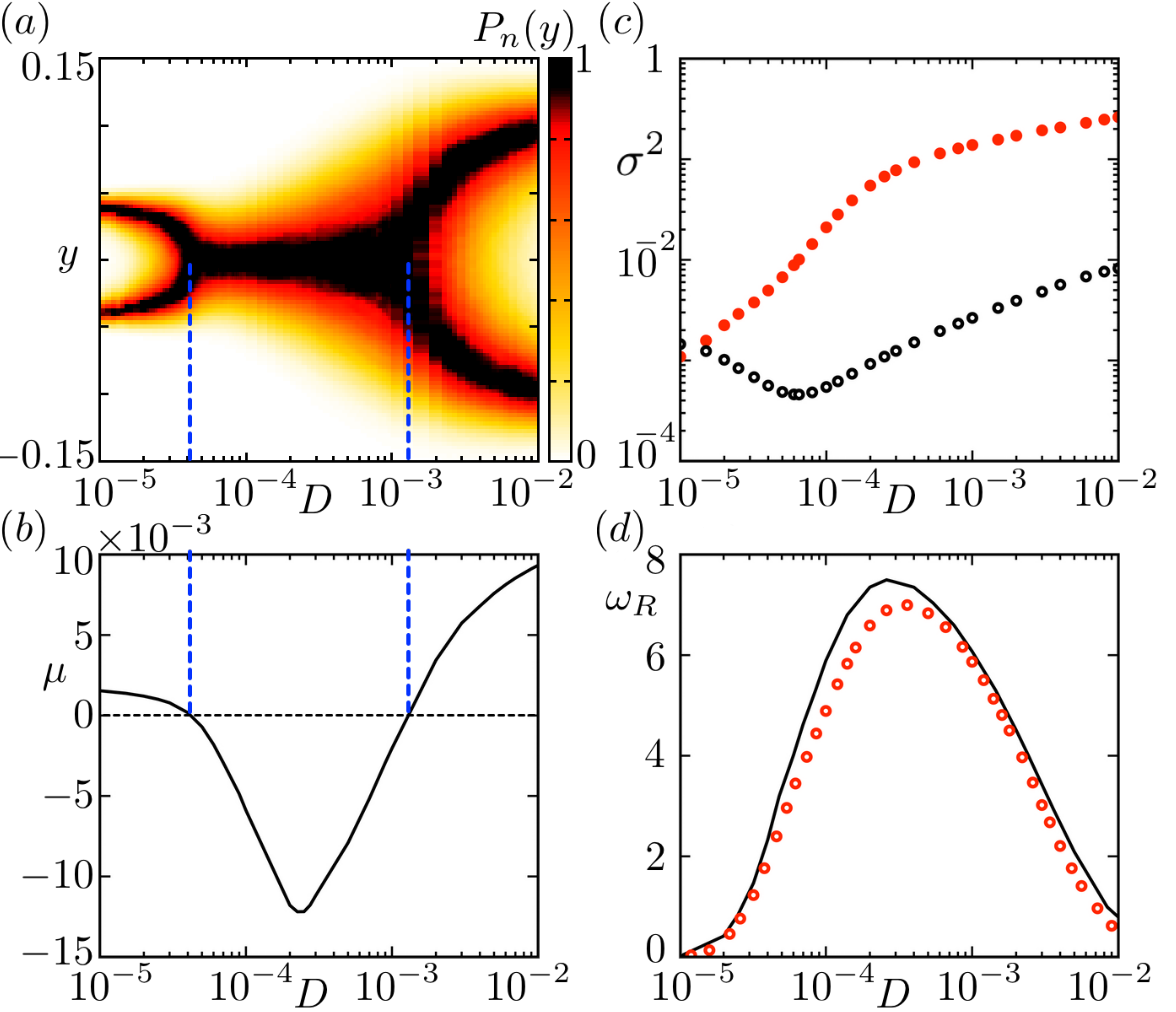} 
\caption{Noise-induced bifurcations in numerical simulations of Eqs.~(\ref{system2}). (a): Normalized coordinate PDF, $P(y)/P_\mathrm{max}$ vs noise intensity.  
(b): Effective bifurcation parameter $\mu(D)$. Vertical dashed lines indicate positions of two pitchfork bifurcations. (c): Variances of coordinate (open black circles), $\sigma_v^2$, and velocity (filled red circles), $\sigma^2_v$ vs noise intensity. 
(d): Rice frequency, $\omega_R$, vs noise intensity. Red circles show values numerically obtained values; solid line shows $\omega_R$ calculated from Eq.~(\ref{rise}) with the effective potential $\Ueff$ and noise intensity, $\Deff$.  Other parameters are: $\eps=0.01, c_{1}=1, c_{3}=9, c_{5}=22, a=1.2, b=100$.}
\label{fig6}
\end{figure}  

\subsection{Numerical simulations}
Numerical simulations confirmed the existence of  multiple noise-induced transitions in the mathematical model of the circuit Eqs.(\ref{system2}) as shown in Fig.~\ref{fig6}. Double-peaked stationary coordinate PDF, $P(y)$, becomes unimodal with the increase of noise intensity. Further increase of noise gives rise to the backward transition from mono- to bistable dynamics with doubly peaked coordinate PDF shown in  Fig.~\ref{fig6}(a). As for analog simulations, noise-induced transitions are well described by the effective bifurcation parameter, $\mu(D)$, shown in Fig.~\ref{fig6}(b). Similar to analog experiment, the velocity variance, $\sigma^2_v$, which determines the effective noise intensity, shows monotonous increase with $D$, while the coordinate variance possesses a minimum, reflecting the first noise-induced transition from bi- to unimodal structure of the coordinate PDF [compare Figs.~\ref{fig5}(c) and \ref{fig6}(c)]. Similarly, the Rice frequency exhibit non-monotonous dependence on noise intensity attaining its maximum  at $D$ corresponding to the minimal value of effective bifurcation parameter, $\mu(D)$ [compare Figs.~\ref{fig5}(d) and \ref{fig6}(d)].

\section{Mechanism of noise-induced transitions} 
\begin{figure}[t]
\centering
\includegraphics[width=0.5\textwidth]{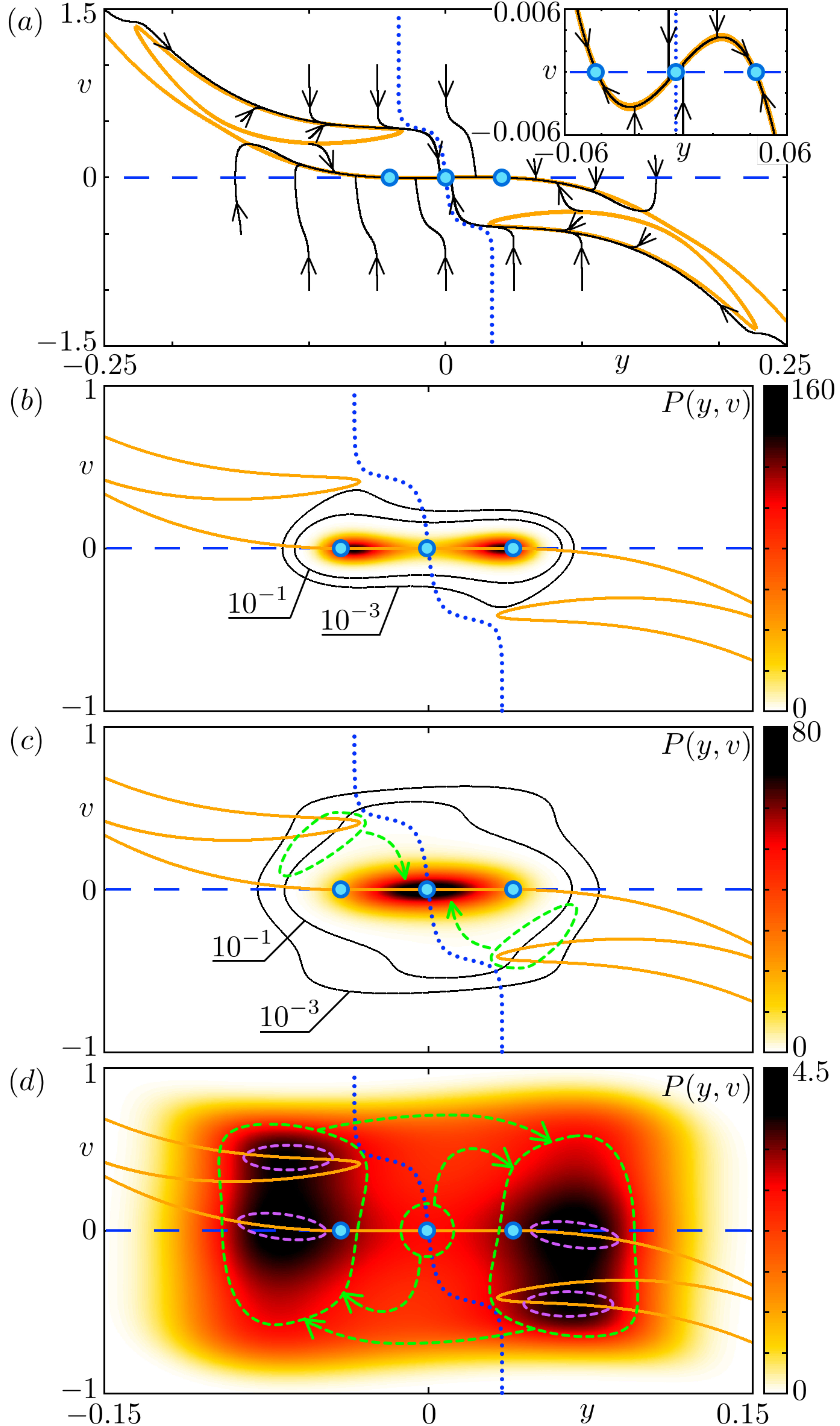} 
\caption{Mechanism of noise-induced pitchfork bifurcations. On all panels: equilibrium points are shown by blue circles; blue dashed line indicates the nullcline $\dot{y}=0$; orange solid line shows the nullcline $\dot{v}=0$; the separatrix of the saddle at the origin is shown by blue dotted line.
(a): Deterministic dynamics of 
Eq.(\ref{system2}). Phase trajectories started from various initial conditions are shown by black arrowed lines. Inset shows expanded region near equilibria.
Panels (b)--(d) also show  contour maps of the stationary PDF, $P(y,v)$, obtained numerically. 
(b): $D=2\times 10^{-5}$; (c): $D=6\times 10^{-5}$;
(d): $D=2.4 \times 10^{-3}$. Other parameters are: 
 $\eps=0.01, c_{1}=1, c_{3}=9, c_{5}=22, a=1.2, b=100$, i.e. the same as in analog experiments.}
\label{fig7}
\end{figure}

Noise-induced changes in the oscillator's dynamics
can be understood by studying the structure of the phase space of the corresping deterministic system described by Eqs.(\ref{system2}) with $D=0$. Fig.~\ref{fig7}(a) shows two stable nodes, separated by the saddle at the  origin, and nullclines of the system. The intrinsic feature of the system under study is an unusual structure of the nullcline $\dot{v}=0$. Besides the conventional N-shape branch passing through equilibria [inset in Fig.~\ref{fig7}(a)], the nullcline includes two symmetric separate closed-loop branches.  Let us consider a loop at the upper left quadrant in Fig.~\ref{fig7}(a).
The upper side of the loop is attractive and the lower side  is repulsive. When a phase trajectory approaches the loop from above, it slows down and moves on the attractive side until it  approaches the separatrix of the saddle, and  then eventually falls onto  vicinity of the saddle equilibrium at the origin. Repulsive side of the close-loop branch directs phase trajectories towards the stable equilibrium. Symmetrical behavior occurs for the loop at the right lower quadrant.  We note, that for the linear resistor N1 the nullcline $\dot{v}=0$ has a single N-shape branch only, i.e.  no closed-loop segments.

Weak noise  results in conventional stochastic hoping between two metastable states with the double-peaked PDF [Fig.~\ref{fig7}(b)]. Note, that probability to cross closed-loop branches of the nullcline is rather low and so they have no effect on the position of the PDF's maxima.
With the increase of noise intensity the PDF $P(y,v)$ smears vertically, i.e. with respect to velocity, $v$. Fig.~\ref{fig7}(c) indicates, that phase trajectories frequently visit closed-loop branches 
[areas are marked by the green dashed line on Fig. \ref{fig7}(c)] which results in deflection towards the origin, as described for deterministic case on Fig. \ref{fig7}(a). This results in shifting of the PDF's maxima towards the origin. There is a critical noise intensity which corresponds to maximal influence of the closed-loop nullcline branches, resulting in the PDF  with single peak at the origin. 
For larger values of noise intensity phase trajectories  begin to pass through the repulsive sides of the closed-loop nullcline branches. They are then slow down on attractive branches of the nullcline [areas marked by violet dashed line on Fig.~\ref{fig7}(d)] and then deflected towards the origin. However, because of larger noise, phase trajectories can now overcome the separatrix towards another stable equilibrium, rather than fall onto the origin. As a result, the origin is visited less frequently than symmetrical areas on the left and right of the separatrix. In this way the central peak of the PDF becomes divided into two and bistability is restored.

Analysis of the oscillator model has shown that S-shaped I-V characteristic of  nonlinear element N1 (Fig.\ref{fig1}) with fifth-order nonlinearity in the function $F(x)$ in Eq.~(\ref{initial}) is necessary for specific nullclines arrangements shown in Fig.~\ref{fig7}(a) and so  for the observed noise-induced transitions. Addition of higher order nonlinearities (e.g. seventh-order term) to $F(x)$ does not change qualitatively the dynamics of the system.

\section{Conclusions} We have developed a generic model of the bistable oscillator with nonlinear dissipation. Using analog circuit experiment and numerical simulation we showed that this system demonstrates multiple noise-induced transitions, registered as changes of extrema in the stationary PDF. Using the effective potential approach we showed that the observed noise-induced transitions are described by a normal form of pitchfork bifurcation.  
We note that qualitative structural changes in the stationary PDF  are not {\it pure} noise-induced transitions \cite{horsthemke1984}, as the bistability is inherent for the model. Nevertheless, the observed transitions from bimodal and unimodal and back to bimodal cannot be realized in the system by changing a single parameter of nonlinear or linear components of the circuit. Thus, noise intensity is an independent control parameter which determines qualitative nature of the system's dynamics. 
We provided a clear explanation of the mechanism of the effect based on partition of the phase space of the system by nullclines and manifolds of the saddle equilibrium.

\section*{Acknowledgements} 
This work was supported by the Russian Foundation for Basic Research (RFBR) (grant No. 15-02-02288) and by the Russian Ministry of Education and Science (project code 1008). We are very grateful to E. Sch{\" o}ll, A. Zakharova, and G. Strelkova for helpful discussions. ABN gratefully acknowledges the support of NVIDIA Corp. with the donation of the Tesla K40 GPU used for this research.


%

\end{document}